\documentclass[twocolumn,preprintnumbers,amssymb,aps,pra]{revtex4-1}
\usepackage{pgfplots}
\usepackage{latexsym,epsfig}
\usepackage{graphicx}
\usepackage{dcolumn}
\usepackage[normalem]{ulem}
\usepackage{amsthm,amsmath}

\begin{document}
\title{Determination of the dipole polarizability of the alkali-metal negative ions}

\author{B. K. Sahoo}

\email{bijaya@prl.res.in}

\affiliation{Atomic, Molecular and Optical Physics Division, Physical Research Laboratory, Navrangpura, Ahmedabad-380009, India}

\date{Received date; Accepted date}

\begin{abstract}
We present electric dipole polarizabilities ($\alpha_d$) of the alkali-metal negative ions, from H$^-$ to Fr$^-$, by employing four-component
relativistic many-body methods. Differences in the results are shown by considering Dirac-Coulomb (DC) Hamiltonian, DC Hamiltonian 
with the Breit interaction, and DC Hamiltonian with the lower-order quantum electrodynamics interactions. At first, these interactions 
are included self-consistently in the Dirac-Hartree-Fock (DHF) method, and then electron correlation effects are incorporated over
the DHF wave functions in the second-order many-body perturbation theory, random phase approximation and coupled-cluster (CC) 
theory. Roles of electron correlation effects and relativistic corrections are analyzed using the above many-body methods with size 
of the ions. We finally quote precise values of $\alpha_d$ of the above negative ions by estimating uncertainties to the CC results, 
and compare them with other calculations wherever available. 
\end{abstract}

\maketitle

There are several experimental techniques available to produce negative alkali ions in the laboratory. The techniques to produce these ions have 
been engineered time to time over the several decades \cite{vermeer,andersen1}. Electron affinities (EAs) of these systems have been measured very 
precisely \cite{hotop,patterson,kasdan,andersen2,andersson}. Photoabsorption spectra of these ions have also been extensively investigated both 
theoretically and experimentally, in order to understand their structures \cite{watanabe,eiles,berzinsh,lindhal}. Starting from seventies, a series
of studies on the photodetachment of negative lithium (Li$^-$), sodium (Na$^-$) and potassium (K$^-$) ions have been conducted by several groups
\cite{kasdan,haeffler,moores,slater,taylor}. Theoretical results from these lighter ions were in excellent agreement with the corresponding 
experimental values. By solving a set of coupled equations, Norcross had predicted the existence of bound excited states in negative alkali ions 
\cite{norcross}. This was later conformed by Greene \cite{greene}, by applying a combined approach of $jj$-coupling in the $R$-matrix method and 
generalized quantum-defect theory while attempting to describe photodetachment spectra of negative rubidium (Rb$^-$), cesium (Cs$^-$) and francium 
(Fr$^-$) ions. Later, another study \cite{bahrim} disproved the existence of such states in these ions by reanalyzing the calculations using the Dirac
$R$-matrix method. Instead, the authors of the work suggested that the lowest excited state of the above alkali negative ions is a multiplet 
of $^3 P_J^o$-shape resonance. This work clearly demonstrated the importance of relativistic effects for accurate calculations of atomic properties in 
these ions. Similarly, several studies on negative hydrogen (H$^-$) ion have been carried out \cite{balling,wildt}, and its applications 
\cite{millar,rau} and production sources are well-known to the physicists \cite{dudinikov,faircloth}. 

Apart from EAs and photodetachment cross-sections, there is scarcity in the atomic data of the negative alkali ions. Alkali atoms have a closed-shell
and a valence orbital in their electronic configurations. Owing to this, it is relatively simpler to calculate atomic wave functions of these atoms. 
However, it is challenging to determine atomic wave functions of the alkaline earth-metal atoms due to strong electron correlations among the valence
electrons in such systems. The negative alkali ions are isoelectronic to the alkaline earth-metal atoms. 

Electric dipole polarizability ($\alpha_d$) is a very useful property of any atomic system. This quantity has been measured very precisely 
\cite{schwerdtfeger2} in the alkali atoms as well as in the singly charged positive alkaline earth-metal ions, and with reasonable accuracy in the 
alkaline earth-metal atoms. However, there has not been a single measurement of $\alpha_d$ carried out thus far in any of the negative alkali ions 
due to difficulties in setting up their experiments. There are no full relativistic calculation of $\alpha_d$ available in these ions, and only a 
few non-relativistic calculations have been reported in the lighter H$^-$ \cite{pipin,kar,bhatia}, Li$^-$ \cite{moores,pouchan,agren}, 
Na$^-$ \cite{moores} and K$^-$\cite{moores} ions. Except the high-precision calculations in H$^-$, the reported values of other ions are not 
very reliable. 

Calculations of $\alpha_d$ in the alkali atoms are in very good agreement with the experiments \cite{arora1,arora2}. The reason for this is that one
can easily use their experimental energies and electric dipole (E1) matrix elements inferring from the lifetime measurements of their atomic states
in the evaluation of $\alpha_d$ values using the sum-over-states approach. It is possible to adopt the sum-over-states approach in these atoms because they possess 
a large number of bound states in contrast to the negative alkali ions. {\it Ab initio} procedures demonstrate that the electron correlation and the 
relativistic effects are pronounced, and they need to be accounted for, in order to determine $\alpha_d$ values of alkaline earth-metal atoms 
\cite{lim,bijaya1,yashpal1}, which are isoelectronic systems to the negative alkali ions. A few calculations of the $\alpha_d$ values of some heavier
negative ions of the coinage metal atoms have been reported by Sadlej and coworkers \cite{diercksen,kello}, by employing a variety of methods 
including the coupled-cluster (CC) theory. In another study by Schwerdtfeger and Bowmaker \cite{schwerdtfeger}, these quantities were also evaluated 
by using total angular momentum $j$-averaged relativistic pseudo-potentials in the configuration interaction (CI) method. These works highlight about the 
unusually large electron correlation and relativistic effects in the determination of $\alpha_d$ values in the negative ions compared to their counter
isoelectronic neutral atoms. But the relativistic effects were estimated only approximately in these calculations. Recently, we had evaluated 
$\alpha_d$ values of Cl$^-$ and Au$^-$ by applying a number of relativistic many-body methods at different levels of approximation to demonstrate the
roles of electron correlations for their accurate determination. Here, we intend to determine $\alpha_d$ values of all the negative alkali ions 
very accurately.

It is not possible to adopt a finite-field (FF) approach to determine $\alpha_d$ of atomic states by preserving spherical symmetry property and 
treating parity as a good quantum number. Thus, the spherical symmetry of the systems is exploited in Refs. \cite{diercksen,kello,schwerdtfeger}
in order to adopt the FF procedure for the determination of $\alpha_d$ of negative ions. Also, the FF approach introduces large uncertainty to the 
calculation of $\alpha_d$, which stems from both numerical differentiation as well as neglecting higher-order perturbation corrections in the 
evaluation of the second-order perturbed energy due to external electric-field, in a brute-force manner. To overcome these problems in the 
determination of $\alpha_d$ while retaining spherical symmetry behavior of atomic orbitals, we follow a perturbative approach in which the total
Hamiltonian of the system is defined in the presence of a weak external electric field $\vec{\mathcal E}$ as $H=H_{at}+\vec D \cdot \vec{\mathcal
E}$ with the atomic Hamiltonian $H_{at}$ and electric dipole operator $D$ in a similar framework as Dalgarno and Lewis \cite{dalgarno}. In such case, the wave function and energy of an atomic state can be expressed as
\begin{eqnarray}
 && |\Psi_0 \rangle = |\Psi_0^{(0)} \rangle + |\vec{\mathcal E}| |\Psi_0^{(1)} \rangle + \cdots  \label{eq1} \\
 \text{and} && \nonumber \\ 
&& E_0 = E_0^{(0)} + |\vec{\mathcal E}| E_0^{(1)} +  \frac{1}{2} |\vec{\mathcal E}|^2 E_0^{(2)} \cdots , \label{eq2} 
\end{eqnarray}
respectively, where superscripts (0), (1), etc. denote order of $\vec{\mathcal E}$ in the expansion. Since $D$ is an odd-parity operator, $E_0^{(1)}=0$ 
and the second-order energy is traditionally given by $E_0^{(2)} \equiv \alpha_d$. It follows that $\alpha_d$ can be evaluated in the perturbative 
approach as \cite{bijaya1}
\begin{eqnarray}
\alpha_d &=&  2 \frac{\langle \Psi_0^{(0)}|D|\Psi_0^{(1)} \rangle}{ \langle \Psi_0^{(0)}| \Psi_0^{(0)} \rangle } . \label{eq3}
\end{eqnarray}
Thus, it is imperative to determine both the unperturbed wave function $|\Psi_0^{(0)} \rangle$ of $H_{at}$ and the first-order perturbed wave function 
$|\Psi_0^{(1)} \rangle$ due to $D$ very reliably for an accurate evaluation of $\alpha_d$. Instead of using the sum-over-states approach to determine 
$|\Psi_0^{(1)} \rangle$, we would like to solve it as the solution to the first-order inhomogeneous perturbed equation given by 
\begin{eqnarray}
 (H_{at}-E_0^{(0)}) |\Psi_0^{(1)} \rangle = - D  |\Psi_0^{(0)} \rangle .  \label{eq4}
\end{eqnarray}
Though the solution of this equation appears to be similar to the procedure adopted by Dalgarno and Lewis \cite{dalgarno}, but it can be kept in 
mind that we only obtain the first-order wave function $ |\Psi_0^{(1)} \rangle$ for Eq. (\ref{eq3}) instead of determining $E_0^{(2)}$ directly.

The many-electron atomic wave function can be obtained by
\begin{eqnarray}
|\Psi_0 \rangle = \Omega_0 |\Phi_0 \rangle , \label{eq5}
\end{eqnarray}
where $|\Phi_0 \rangle$ is a mean-field wave function, which is obtained here by the Dirac-Hartree-Fock (DHF) method, and $\Omega_0$ is known as 
the wave operator that is responsible for accounting for electron correlation effects due to the interactions that are neglected in the determination 
of $|\Phi_0 \rangle$. Likewise, for the wave function, we can expand $\Omega_0$ in the presence of weak electric field $\vec{\mathcal E}$ as
\begin{eqnarray}
\Omega_0 = \Omega_0^{(0)} + |\vec{\mathcal E}|\Omega_0^{(1)}  + \cdots . \label{eq6}
\end{eqnarray}
Using this, we can write the unperturbed and the first-order perturbed wave function as  
\begin{eqnarray}
 |\Psi_0^{(0)} \rangle = \Omega_0^{(0)} |\Phi_0 \rangle  \ \ \ \text{and} \ \ \
 |\Psi_0^{(1)} \rangle = \Omega_0^{(1)} |\Phi_0 \rangle  . \label{eq7}
\end{eqnarray}

In the $n^{th}$-order perturbation theory, we express \cite{yashpal1}
\begin{eqnarray}
 \Omega^{(0)} &=& \sum_{k=0}^n \Omega^{(k,0)} \ \ \ \text{and} \ \ \ 
 \Omega^{(1)} = \sum_{k=0}^{n-1} \Omega^{(k,1)}
\end{eqnarray}
with $\Omega^{(0,0)}=1$, $\Omega^{(1,0)}=0$ and $\Omega^{(0,1)}= \sum_{p,a} \frac{ \langle \Phi_a^p | D | \Phi_0 \rangle} {\epsilon_a^{(0)} - 
\epsilon_p^{(0)}}$ for all the occupied orbitals denoted by the index $a$ and unoccupied orbitals denoted by the index $p$. In the second-order
relativistic perturbation theory (RMBPT(2) method) that accounts for the lowest-order electron correlation effects \cite{yashpal1} in the many-body 
theory, it corresponds to $n=1$ in the above summations. 

We present results from two all-order many-body methods: relativistic random-phase approximation (RRPA) and relativistic CC (RCC) theory. The RCC 
theory incorporates electron correlation effects more rigorously, while RRPA has traditionally been employed to capture these effects due to the 
core-polarization only, which can be done to all-orders in a computationally much less expensive way. The correlation effects arising through 
RRPA also represent the orbital relaxation effects that arise naturally in the mixed-parity orbitals of the DHF method in the FF procedure. 
In our RRPA implementation \cite{yashpal2,bijaya2}, they are contained in $\Omega_0^{(0)}=1$ and $\Omega^{(1)} =  \sum_{k=0}^{\infty} \sum_{p,a}
\Omega_{a \rightarrow p}^{(k, 1)}$. Here, $a \rightarrow p$ means replacement of an occupied orbital $a$ from $|\Phi_0 \rangle$ by a virtual 
orbital $p$, which alternatively refers to a singly excited state with respect to $|\Phi_0 \rangle$. The RCC theory implicitly includes 
correlation effects arising through RRPA along with other correlation effects such as pair-correlation effects to all-orders and is known as the 
gold standard method of many-body theory for its capabilities of producing accurate results in multi-electron systems. In this theory, the wave 
operators are given by \cite{yashpal2,bijaya3}
\begin{eqnarray}
\Omega^{(0)} &=& e^{T^{(0)}}  \ \ \ \ \text{and} \ \ \ \ \Omega^{(1)}= e^{T^{(0)}} T^{(1)} ,
\end{eqnarray} 
respectively. We consider only singles and doubles excitations in the RCC calculations (RCCSD method) by expressing 
\begin{eqnarray}
T^{(0)} = T_1^{(0)}+ T_2^{(0)} \ \ \ \ \text{and}  \ \ \ \ T^{(1)} = T_1^{(1)}+ T_2^{(1)},
\end{eqnarray} 
where subscripts (1) and (2) denote the level of excitation. In this method,  $\alpha_d$ determined as
\begin{eqnarray}
 \alpha_d &=& 2 \frac{\langle\Phi_0 | \Omega^{(0) \dagger} D \Omega^{(1)} | \Phi_0 \rangle }
                  {\langle\Phi_0 | \Omega^{(0) \dagger} \Omega^{(0)} | \Phi_0 \rangle } 
        = 2 \langle\Phi_0 |(\overbrace{D^{(0)}} T^{(1)})_c|\Phi_0 \rangle, \ \ \ \ \
\end{eqnarray}
where $\overbrace{D^{(0)}} = e^{T^{\dagger{(0)}}}De^{T^{(0)}}$ is a non-truncating series. We have adopted an iterative procedure to take into 
account contributions from this non-terminating series self-consistently, as described in our earlier works on $\alpha_d$ calculations 
in the closed-shell atoms \cite{yashpal3,bijaya3}.

\begin{table}[t]
\caption{Calculated $\alpha_d$ values (in $ea_0^3$) of the negative alkali-metal ions from the DHF, RMBPT(2), RRPA and RCCSD methods. Results from 
the DC, DCB and DCQ Hamiltonians are listed separately to highlight the roles of Breit and QED interactions in the determination of $\alpha_d$ of 
the above ions.}
\begin{ruledtabular}
\begin{tabular}{lccc} 
   Method    & DC         &   DCB     & DCQ \\
 \hline \\
 \multicolumn{4}{c}{H$^-$ ion} \\
 DHF         &    44.41   &    44.61  &   44.41 \\
 RMBPT(2)    &    66.35   &    66.35  &   66.35 \\
 RRPA        &    91.13   &    91.13  &   91.13 \\
 RCCSD       &   206.14   &   206.16  &  206.15 \\
 \hline \\
  \multicolumn{4}{c}{Li$^-$ ion} \\
 DHF         &   500.90   &   500.94  &   500.91 \\
 RMBPT(2)    &   764.68   &   764.74  &   764.70 \\
 RRPA        &  1176.68   &  1176.76  &  1176.70 \\
 RCCSD       &   794.06   &   794.08  &   794.07 \\
 \hline \\
  \multicolumn{4}{c}{Na$^-$ ion} \\
 DHF         &   605.91   &   606.00     606.03 \\
 RMBPT(2)    &   923.52   &   923.66  &  923.72  \\
 RRPA        &  1447.69   &  1447.92  & 1448.01 \\
 RCCSD       &   952.54   &   952.63  &  952.67 \\
 \hline \\
  \multicolumn{4}{c}{K$^-$ ion} \\
 DHF         &  1053.39  &  1053.58   & 1053.89 \\
 RMBPT(2)    &  1586.91  &  1587.18   & 1587.68 \\
 RRPA        &  2565.41  &  2565.88   & 2566.66 \\
 RCCSD       &  1353.69  & 1353.84    & 1354.19 \\
 \hline \\
  \multicolumn{4}{c}{Rb$^-$ ion} \\
 DHF         &  1214.16  &  1214.43   &  1215.40  \\
 RMBPT(2)    &  1816.88  &  1817.22   &  1818.75 \\
 RRPA        &  2968.55  &  2969.21   &  2971.66 \\
 RCCSD       &  1506.57  &  1506.75   &  1507.88  \\
 \hline \\
  \multicolumn{4}{c}{Cs$^-$ ion} \\
 DHF         &  1534.15  &  1534.48  &  1537.06 \\
 RMBPT(2)    &  2271.42  &  2271.78  &  2275.81 \\
 RRPA        &  3770.88  &  3771.71  &  3778.25 \\
 RCCSD       &  1800.42  &  1800.54  &  1803.59 \\
 \hline \\
  \multicolumn{4}{c}{Fr$^-$ ion} \\
 DHF         &   1357.23 &  1357.78  & 1357.33 \\
 RMBPT(2)    &   1990.66 &  1991.21  & 1990.77 \\
 RRPA        &   3308.49 &  3309.76  & 3308.69\\
 RCCSD       &   1619.04 &  1619.16  & 1619.03\\
\end{tabular}
\end{ruledtabular}
\label{tab1}
\end{table}

For the evaluation of $|\Psi_0 \rangle$, we consider first the Dirac-Coulomb (DC) Hamiltonian, given by
\begin{eqnarray}\label{eq:DHB}
H^{DC} &=& \sum_i \left [c\mbox{\boldmath$\alpha$}_i\cdot \textbf{p}_i+(\beta_i-1)c^2+V_n(r_i)\right] +\sum_{i,j>i}\frac{1}{r_{ij}}, \ \ \ \ \ \
\end{eqnarray}
where $c$ is the speed of light, $\mbox{\boldmath$\alpha$}$ and $\beta$ are the usual Dirac matrices, $\textbf{p}_i$ is the single particle 
momentum operator, $V_n(r_i)$ denotes the nuclear potential, and $\frac{1}{r_{ij}}$ represents the Coulomb potential between two electrons 
located at the $i^{th}$ and $j^{th}$ positions. We estimate the Breit interaction by using the Dirac-Coulomb-Breit (DCB) Hamiltonian 
($H^{DCB}=H^{DC}+V^B$) by defining the potential
\begin{eqnarray}\label{eq:DHB}
V^B &=& - \sum_{j>i}\frac{[\mbox{\boldmath$\alpha$}_i\cdot\mbox{\boldmath$\alpha$}_j+
(\mbox{\boldmath$\alpha$}_i\cdot\mathbf{\hat{r}_{ij}})(\mbox{\boldmath$\alpha$}_j\cdot\mathbf{\hat{r}_{ij}})]}{2r_{ij}} ,
\end{eqnarray}
where $\mathbf{\hat{r}_{ij}}$ is the unit vector along $\mathbf{r_{ij}}$. Similarly, contributions from the quantum electrodynamics (QED) effects
are estimated using the Dirac-Coulomb-QED (DCQ) Hamiltonian ($H^{DCQ}=H^{DC}+V^Q$) by considering $V^Q=V_{VP}+V_{SE}$ with the vacuum polarization 
interaction potential $V_{VP}$ and the self-energy interaction potential $V_{SE}$. We use the model potentials for $V_{VP}$ and $V_{SE}$ as 
defined in Refs. \cite{ginges,ymyu}.

We use Gaussian type orbitals (GTOs), as defined in Ref. \cite{mohanty}, to obtain the single particle orbitals. We have considered orbitals
up to $h$-angular momentum symmetry (orbital angular momentum $l=5$) to carry out all the calculations. We have used 40 GTOs for each symmetry to 
obtain the DHF wave function. However, we have frozen high-lying orbitals beyond energy 3000 atomic units (a.u.) to account for electron 
correlation effects through the employed many-body methods. We have verified contributions from these neglected orbitals using RRPA and they are 
found to be extremely small. These contributions are included in the uncertainty estimation later.

\begin{figure}[t]
%centering
\includegraphics[width=8.5cm,height=5cm]{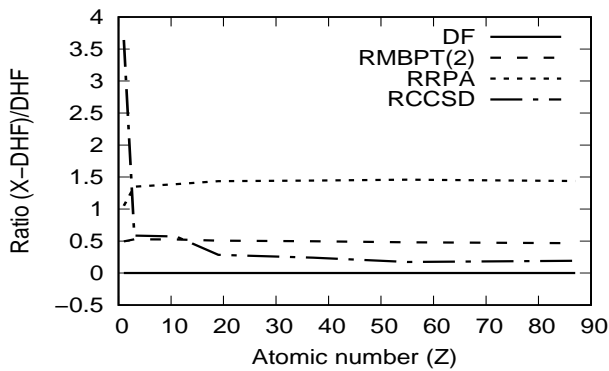}
\caption{\label{fig1} (X-DHF)/DHF values, where X represents contributions from the DHF, RMBPT(2), RRPA and RCCSD methods, against the atomic number 
($Z$) of the negative alkali ions. Values from the DHF method act as reference (zero line on x-axis) for comparison of correlation contributions 
incorporated at different levels of many-body methods.}
\end{figure}

 In Table \ref{tab1}, we present $\alpha_d$ values of all the negative alkali ions in a.u., $ea_0^3$, from the DHF, RMBPT(2), RRPA and RCCSD 
methods using the DC, DCB and DCQ Hamiltonians. It can be seen from this table that the DHF method gives lower values in all the ions, and the 
electron correlation contributions enhance their magnitudes. Except in the H$^-$ ion, the RRPA yields the largest value for each ion. The results from the RMBPT(2)
method are also larger than the values obtained using the RCCSD method, except in the lighter H$^-$ and Li$^-$ ions. It can be recalled that the 
results from the RMBPT(2) method are actually the lowest-order core-polarization terms and results from RRPA are the contributions from 
the all-order core-polarization effects including the DHF value. Thus, the huge differences between the results from the RMBPT(2) and RRPA methods 
imply that the core-polarization effects arising through the higher-order perturbation theory are quite strong in the negative alkali ions. Since 
RRPA values are the mean-field contributions (i.e. DHF values) in the FF procedure, it would require immense efforts to attain convergence in the 
values by evaluating the property in the FF framework than in the perturbation approach through a many-body method as adopted here. Nonetheless, the RCCSD method 
implicitly incorporates the RRPA contributions in addition to correlation effects due to non-RRPA effects such as those from the pair-correlations. Large
differences between the RRPA and RCCSD results indicate that the non-RRPA contributions are also substantially large, but with opposite sign than that
of the RRPA contributions. As a result, the final results in the RCCSD method come out to be smaller than the RMBPT(2) values in the heavier ions. 

The electron correlation effects in the simplest two electron H$^-$ ion, which is analogous to He atom, shows a unique trend than the rest of the 
ions. In this system, the inclusion of electron correlation effects increases the $\alpha_d$ values gradually through the RMBPT(2), RRPA and RCCSD methods. 
Recently, we have evaluated this property for the negative chlorine (Cl$^-$) and gold (Au$^-$) ions by employing the above many-body methods 
\cite{bijaya2}. We find similar trends in the electron correlation effects in H$^-$ and Cl$^-$, but the correlation contributions at different levels of 
approximations in the many-body theory are found to be quite large in H$^-$ compared to Cl$^-$. The DHF value of $\alpha_d$ was larger than the RCCSD 
value in Au$^-$, in contrast to the trend seen in the negative alkali ions. In Fig. \ref{fig1}, we plot the fractional differences of the results from the 
DC Hamiltonian from different methods against the atomic number ($Z$) of the negative alkali ions. This shows that scaling of correlation contributions 
through RMBPT(2) and RRPA varies linearly with the size of  the considered systems, but no particular trend is followed by the contributions from the RCCSD method.

\begin{table}[t]
\scriptsize
\caption{Contributions from the DC Hamiltonian through different RCC terms in the determination of $\alpha_d$ (in $ea_0^3$) of the negative alkali 
ions. The differences between the sum of the contributions from the mentioned terms and the final values from the RCCSD method given using the DC 
Hamiltonian in Table \ref{tab1} correspond to the contributions from the remaining RCC terms that are not shown explicitly here.}
\begin{ruledtabular}
\begin{tabular}{lccccc}
 Ion    &  $DT_1^{(1)}$ & $T_1^{(0)\dagger}DT_1^{(1)}$ &  $T_2^{(0)\dagger}DT_1^{(1)}$ & $T_1^{(0)\dagger}DT_2^{(1)}$ &  $T_2^{(0)\dagger}DT_2^{(1)}$ \\
\hline \\
H$^-$   &     149.59    &                 30.47        &            $-3.88$            &  $-1.99$     &  32.25 \\
Li$^-$  &     968.86    &   $-146.32$                  &        $-108.09$              &   25.45      & 127.48 \\
Na$^-$  &    1160.42    &   $-193.82$                  &        $-118.33$              &   34.24      & 133.65 \\
K$^-$   &    1741.75    &   $-358.37$                  &        $-204.67$              &   69.89      & 197.74  \\
Rb$^-$  &    1948.21    &   $-414.85$                  &        $-224.50$              &   82.39      & 209.90 \\
Cs$^-$  &    2351.13    &   $-516.10$                  &        $-278.24$              &  105.36      & 250.74 \\
Fr$^-$  &    2076.99    &   $-448.47$                  &        $-214.02$              &   87.29      & 190.49 \\
\end{tabular}
\end{ruledtabular}
\label{tab2}
\end{table}

We notice from Table \ref{tab1} that both the higher-order relativistic effects due to the Breit and QED interactions do not contribute 
significantly to $\alpha_d$ of the considered ions. However, the trends in the results from different approximations in  
many-body theories demonstrate that the estimated corrections from the Breit and QED interactions vary in different methods. It is, therefore, not 
prudent to estimate corrections due to these relativistic effects by applying lower-order many-body methods as done in literature quite often. 

\begin{figure}[t]
%centering
\includegraphics[width=8.5cm,height=5cm]{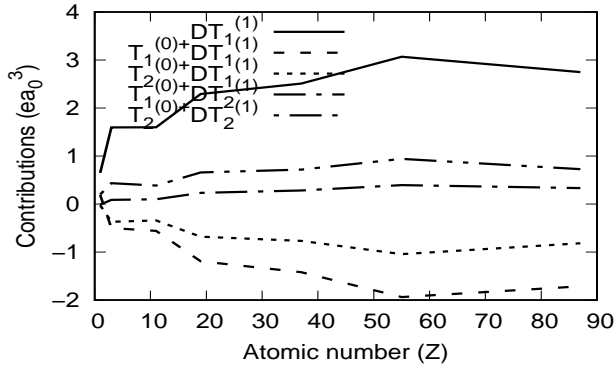}
\caption{\label{fig2} Plot demonstrating correlation contributions arising through the leading order RCC terms against $Z$ of the negative alkali 
ions. The DHF value is subtracted from the first term to show only its correlation contribution.}
\end{figure}

To gain a deeper insight into the behavior of electron correlation effects in the determination of $\alpha_d$ of the negative alkali ions, we 
present contributions from different RCC terms in Table \ref{tab2} by using the DC Hamiltonian. Contributions from the higher-order non-linear RCC 
terms are not given explicitly, but their importance can be found from the differences in the results after summing contributions from the shown 
terms and the final RCCSD values from the DC Hamiltonian given in Table \ref{tab1}. Contrasting behaviors of correlation effects between H$^-$ and 
other ions can be visibly noticeable. The first term, $DT_1^{(1)}$, includes the DHF value, and the leading-order RRPA and non-RRPA correlation contributions 
\cite{bijaya1}. This is the reason for the seemingly dominant contribution of the term over all others. Though the overall trend of correlation effects between 
H$^-$ and Cl$^-$ was earlier found to be similar, comparison of individual contributions from various RCC terms of the above table and that 
given in Ref. \cite{bijaya2} for Cl$^-$ does not suggest the same. In Cl$^-$ ion, the $DT_1^{(1)}$ term accounts for almost all of the $\alpha_d$ 
value in the RCCSD method, whereas in H$^-$, almost all the RCC terms are found to be significant. Also, it can be found from Ref. 
\cite{bijaya2} that only the first three terms contribute mostly to the determination of $\alpha_d$ in the heavier Au$^-$ ion, where contributions 
from the $T_2^{(0)\dagger}DT_1^{(1)}$ RCC term are found to be quite large in the alkali negative ions. We plot contributions from the above 
RCC terms in Fig. \ref{fig2} to demonstrate their roles quantitatively in different ions. For this purpose, we have plotted $DT_1^{(1)}$ value after
subtracting the DHF result. As seen, the magnitudes of the correlation effects are slowly increasing with the size of the system through each RCC
term, with the only exception being the trend from Cs$^-$ to Fr$^-$.

  In Table \ref{tab3}, we quote the final values along with uncertainties by considering the RCCSD results from the DC, DCB and DCQ Hamiltonians of 
our calculations. The uncertainties are estimated by extrapolating contributions from the finite-size basis functions that were used and the neglected 
higher-level excitations (especially from the triple excitations). Our final values are compared with the available literature results in the same 
table. It is worth noting again that due to complication in measuring $\alpha_d$ of the negative ions, there are no experimental values available till 
date and calculations are reported only for a few lighter negative ions by employing lower-order non-relativistic many-body methods. We find that 
among the negative alkali ions, precise results are available only for H$^-$, by applying few-body methods. Though the RCC theory is not apt to 
investigate electron correlation effects in H$^-$, the applied RCCSD method in this work is the complete {\it ansatz} of the RCC theory for this ion. Thus, its 
result can serve as the benchmark to our calculations for other ions. A very good agreement of our result with the reported precise value \cite{bhatia} implies reliability in the calculation using the RCCSD method. In this context, we would like to emphasize here that the DHF, RMBPT(2) and 
RRPA results are far-off from the RCCSD result in this ion. The entire uncertainty to its $\alpha_d$ value arises due to construction of basis
functions owing to its point-size nucleus in contrast to other ions. An old theoretical study presents $\alpha_d$ values of Li$^-$, Na$^-$ and K$^-$ 
by analyzing photodetachment cross-sections in both the length and velocity gauge approximations \cite{moores}. Values from both the gauge expressions
differ widely, raising questions about their accuracies. Two more calculations for $\alpha_d$ of Li$^-$ using CI methods are found in literature \cite{pouchan,agren}. However, results from both the CI calculations are far apart and much outside of the quoted error bars. Thus, these results cannot be used assuredly for any application of this ion. Our $\alpha_d$ value of Li$^-$ matches well with the velocity gauge result of Ref. \cite{moores} and 
calculation by \AA{}gren {\it et al} \cite{agren}, but this agreement could be coincidental as large disagreements between our calculations, 
performed in length gauge, with the results from gauges of Ref. \cite{moores} are seen in the  Na$^-$ and K$^-$ ions. To our knowledge, there are no 
theoretical studies on $\alpha_d$ of Rb$^-$, Cs$^-$ and Fr$^-$ available in the literature to make comparative analysis with our calculations.

\begin{table}[t]
\caption{The recommended $\alpha_d$ values (in $ea_0^3$) of negative alkali ions from this work. Previous calculations using non-relativistic 
methods are also given for comparison. The estimated uncertainties are quoted in parentheses.}
\begin{ruledtabular}
\begin{tabular}{lcc} 
 Ion      &    This work   & Others \\
 \hline \\
 H$^-$    &  206.2(5)  &  206.165 \cite{pipin}, 206.37683 \cite{kar} \\
          &            &  206.1487618(37) \cite{bhatia} \\
 Li$^-$   &  794(2)    & 832$^a$ \cite{moores}, 798$^b$ \cite{moores}, 650(50) \cite{pouchan}\\
          &            &  798(5) \cite{agren} \\
 Na$^-$   &  953(5)    & 989$^a$ \cite{moores}, 1058$^b$ \cite{moores} \\
 K$^-$    & 1354(7)   & 1805$^a$ \cite{moores}, 1757$^b$ \cite{moores}\\
 Rb$^-$   & 1508(8)    &  \\
 Cs$^-$   & 1804(10)   &  \\
 Fr$^-$   & 1620(10)   & \\
\end{tabular}
\end{ruledtabular}
$^a$ From the length gauge calculation.\\
$^b$ From the velocity gauge calculation.
\label{tab3}
\end{table}

  We would also like to make an analogy among the roles of electron correlation effects played in the determination of $\alpha_d$ values of 
the positively charged ions and neutral atoms belonging to the isoelectronic sequence of the considered negative ions. Comparing the calculations 
in the positively charged ions and alkaline earth-metal atoms from our earlier work \cite{yashpal1}, and that with the results for the undertaken
negative ions reported in this work, we find that the Breit and QED contributions are negligibly small in all these three types of systems belonging 
to the same isoelectronic sequence. The correlation trends are found to be almost similar among these systems, but the final results are found to be
at least an order bigger in the negative ions than their positively charged counterparts and neutral atoms \cite{bijaya1,yashpal1,schwerdtfeger2}. 
This implies that the negative ions are highly polarized and very sensitive to an applied electric field. 
 
 In summary, we have evaluated electric dipole polarizabilities of all the negative alkali ions very accurately. Propagations of electron correlation 
effects are investigated from lower-order to all-order in perturbation in the determination of these quantities by employing a relativistic 
second-order perturbation theory, random phase approximation and coupled-cluster method on the Dirac-Hartree-Fock calculation. Trends of 
correlation effects from these methods and through different terms of the coupled-cluster approach are demonstrated quantitatively. It shows that the 
roles of electron correlation effects follow almost similar trends in all ions except in H$^-$, on which they bestow a completely different trend. 
Random-phase approximation predicts the values by huge magnitudes, while the coupled-cluster method gave moderate values. This clearly suggests 
that there are huge cancellations among the electron correlation effects arising through the pair-correlation and core-polarization effects. It also 
demonstrates that magnitudes of the correlation effects increase with the size of the system with an exception in the trend from Cs$^-$ to Fr$^-$. We have also 
compared electron correlation trends in the determination of the above-mentioned quantities among the negative alkali ions, and their isoelectronic 
positively charged ions and neutral atoms. This comparison shows that electric dipole polarizabilities in the above ions are about an order larger
than the values of their respective positively charged ions and neutral atoms indicating that negative alkali ions are highly polarized. Higher-order 
relativistic corrections due to the Breit and quantum electrodynamics interactions are found to be negligibly small, but their magnitudes are 
observed to be modified with the electron correlation effects. There are only a few existing calculations of the investigated property available for the
lighter ions. Since it is extremely difficult to measure the electric dipole polarizabilities of the negative alkali ions, our estimated precise 
values of these quantities will be very useful to the applications of these ions. Our results can be further improved after including contributions
from the triple excitations. 
 
 We acknowledge use of Vikram-100 HPC cluster of Physical Research Laboratory, Ahmedabad, India, which we used to carry out computations for this work.

\end{document}